\documentclass[12pt]{article}
\usepackage{setspace}
\usepackage{fancyhdr}
\usepackage[margin=1in, headheight=15pt]{geometry}
\usepackage{lipsum}
\usepackage{epigraph}
\usepackage[nottoc,numbib]{tocbibind}

\usepackage{graphics}
\usepackage{tikz}
\usetikzlibrary{calc}
\usetikzlibrary{decorations.pathreplacing}
\usepackage{amsmath,amsfonts,amsthm}
\usepackage{mathrsfs}
\usepackage{cancel}
\usepackage[normalem]{ulem}
\usepackage{siunitx}
\usepackage{float}
\usepackage{enumitem}
\usepackage{colortbl}
\usepackage[font=footnotesize,labelfont=bf]{caption}
\usepackage{subcaption} 
\usepackage{tabularx}
\newcolumntype{L}[2]{>{\hsize=#1\hsize\columncolor{#2}\raggedright\arraybackslash}X}%
\newcolumntype{R}[2]{>{\hsize=#1\hsize\columncolor{#2}\raggedleft\arraybackslash}X}%
\newcolumntype{C}[2]{>{\hsize=#1\hsize\columncolor{#2}\centering\arraybackslash}X}%
\usepackage{multicol}
\usepackage{textcomp}

\usepackage{color}
\definecolor{mygreen}{rgb}{0,0.6,0}
\definecolor{mygray}{rgb}{0.5,0.5,0.5}
\definecolor{mymauve}{rgb}{0.58,0,0.82}
\definecolor{mylightgray}{gray}{.9}
\definecolor{myblue}{rgb}{.28,.24,.55}
\definecolor{mylightblue}{rgb}{0.74, 0.83, 0.9}
\definecolor{mybrightred}{rgb}{1,.13,.32}
\definecolor{mypink}{rgb}{0.96, 0.76, 0.76}
\definecolor{mylightgreen}{rgb}{0.66, 0.89, 0.63}
\usepackage{colorspace}
\definespotcolor{clear}{rgb}{0,0,0,0}

\usepackage{natbib}
\setcitestyle{aysep={}}

\newcommand\textcite[5]{(#1\citealt{#2}#3\citealt{#4}#5)}
\usepackage{hyperref}
\hypersetup{colorlinks=true, citecolor=darkgray, linkcolor=darkgray, urlcolor=darkgray}
\makeatletter
\renewcommand*{\@fnsymbol}[1]{\ifcase#1\or*\else\@alph{\numexpr#1-1\relax}\fi}
\makeatother

\usepackage{lmodern}
\usepackage{listings}
\usepackage[textsize=tiny, backgroundcolor=mypink, linecolor=red]{todonotes}

\newcommand{\mytitle}{Polarization of Climate Politics Results from Partisan Sorting: Evidence from Finnish Twittersphere}
\newcommand{\shorttitle}{Polarization of Climate Politics Results from Partisan Sorting}
\author{Ted Hsuan Yun Chen\thanks{Faculty of Social Sciences, University of Helsinki},\footnotemark[3]  Ali Salloum\thanks{Department of Computer Science, Aalto University}, \\ Antti Gronow\footnotemark[2], Tuomas Yl\"a-Anttila\footnotemark[2], and Mikko Kivel\"a\footnotemark[3]}
\newcommand{\surname}{Chen et al.}

\title{\mytitle\thanks{Corresponding author: Ted Hsuan Yun Chen (ted.hsuanyun.chen@gmail.com). We thank Risto Kunelius, Pertti Vehkalahti, Kevin Reuning, Boyoon Lee, and participants of the 2019 Comparing Climate Change Policy Networks Workshop held at the University of Bern for useful feedback. This research is produced under the ECANET consortium (Echo Chambers, Experts and Activists: Networks of Mediated Political Communication) funded by the Academy of Finland, grant nos. 320780 (TY) and 320781 (MK). Additional funding comes from the Kone Foundation (AG; grant no. 201804137) and the Strategic Research Council, Academy of Finland (TY; grant no. 312710).}}
\date{\today}

\begin{document}
\pagenumbering{roman}
\singlespacing
\maketitle
\thispagestyle{empty}
\begin{abstract}
\noindent Prior research shows that public opinion on climate politics sorts along partisan lines. However, they leave open the question of whether climate politics and other politically salient issues exhibit tendencies for issue alignment, which the political polarization literature identifies as among the most deleterious aspects of polarization. Using a network approach and social media data from the Twitter platform, we study polarization of public opinion toward climate politics and ten other politically salient topics during the 2019 Finnish elections as the emergence of opposing groups in a public forum. We find that while climate politics is not particularly polarized compared to the other topics, it is subject to partisan sorting and issue alignment within the universalist-communitarian dimension of European politics that arose following the growth of right-wing populism. Notably, climate politics is consistently aligned with the immigration issue, and temporal trends indicate that this phenomenon will likely persist.

\vspace{0.5cm}
	

\end{abstract}

\newpage
\tableofcontents

\newpage
\onehalfspacing
\pagenumbering{arabic}
\setcounter{page}{1}

\section{Introduction}
To varying extents, public opinion on climate politics around the world is polarized \citep{mccright2016political, capstick2015international}. Even in countries where overall levels of climate skepticism is low, the supporters and opponents of climate change mitigation policies tend to be antagonized. Numerous studies show that current levels of polarization has undesirable effects on efforts to address the climate change issue, such as constraining climate communication \citep{zhou2016boomerangs} and decreasing support for both public and private environmental spending \citep{johnson2019political,gromet2013political}. Better understanding of the characteristics of climate polarization is the first step to addressing it. However, despite the urgency of the situation, research into what drives polarization of public opinion in this area remains limited.

Existing work on this topic shows that positions on climate politics is divided along party lines \citep{dunlap2016political,farrell2016corporate,guber2013cooling,mccright2011politicization}. This suggests that the polarized public opinion on climate change is driven by partisan sorting, the phenomenon where the public aligns itself more closely with their preferred parties because the increasing polarization of political elites makes it easier to do so \citep{,hetherington2009putting,fiorina2008political}. Beyond the deleterious consequences introduced by the polarization of climate politics in itself, partisan sorting has the additional implication that the issue becomes more easily captured by extreme interests, leading to uneven political representation \citep{baldassarri2008partisans}. 

Most of this research is based on the United States context \citep{capstick2015international}, which in many ways differs from the rest of the world. First, the most salient climate debate in the U.S. is about beliefs toward the existence of anthropogenic climate change \citep{dunlap2010climate} instead of policy-based debates that are more common elsewhere \citep{cann2018does}. Second, the U.S. has a strong two-party system characterized by both ideological and affective polarization along party lines \citep{fiorina2008political,iyengar2019origins}. This deep political divide combined with the emotive nature of the climate issue makes it unsurprising that positions on climate issues and broader party politics are aligned in the U.S. These studies therefore have relatively low generalizability despite the U.S.'s broader importance in global climate politics. Outside of the U.S. context, studies from Australia \citep{tranter2011political}, Norway \citep{aasen2017polarization}, and the United Kingdom \citep{poortinga2011uncertain}, as well as those conducted cross-nationally \citep{sohlberg2017effect,mccright2016political,tranter2015scepticism} similarly find that individuals' outlooks on climate politics align with their partisanship and ideological positions. Still, the research is sparse, covering only a small subset of countries in the world; in a recent review, \citet{mccright2016political} found only 11 non-U.S. based studies on climate polarization. 

Beyond the issue of limited country cases, two important gaps remain in the literature. First, most previous studies rely primarily on self-reported items from surveys, which largely constrains the examination to the individual level. This in turn limits our ability to observe contextual or emergent dynamics governing public opinion polarization \citep{capstick2015international}. Survey-based cohort analysis \citep[eg.][]{johnson2019political} begins to move beyond individuals, but are rare in the literature. Related to this, common measurement errors in surveys \citep[i.e. tendency to choose middle options, see][]{moors2008exploring} will strongly influence inferences about polarization \citep{hetherington2009putting}. Second, to the extent that existing work shows evidence of partisan sorting in public attitudes toward climate politics, they leave open the question of further alignment between climate and other socially salient issues. In the polarization literature, an important question is whether partisan sorting will yield alignment in individuals' positions across issues \citep{baldassarri2008partisans}, leading to entrenched in-group identities that make it difficult to reach agreements despite the supposed flexibility from working in multidimensional policy space \citep{mason2015disrespectfully}.

Given the current state of the field, further work into a broader set of cases, especially using more diverse methodological approaches and with an additional focus on how climate politics relate to other socially salient issues, will provide a better understanding of how climate politics becomes polarized. We address these gaps by studying polarization of climate politics in Finland using social network data from the Twitter platform. Our study makes a number of contributions. 

First, Finland as a case is advantageous on several fronts. Finland has a multiparty political system, and the climate debate in Finland centers on the tradeoff between environmental and socioeconomic concerns \citep{teravainen2010political}, which is more in line with global trends \citep{cann2018does}. These characteristics make Finland different from the U.S. on the important dimensions of climate politics and political institutions, thereby providing a wide range of coverage. Additionally, Finnish politics has traditionally been characterized by low levels of polarization and tendencies toward consensus politics \citep{reiljan2020fear}. Recent trends, however, indicate increasing polarization among political elites, notably in the areas of climate and immigration politics \citep{lonnqvist2020polarization}. Finland's political history therefore makes it a `hard' case where partisan sorting should be low, and because political elites have begun to strategically politicize the climate issue, it is particularly suitable for studying the effect of partisan sorting on climate polarization.

Second, we take a system-level examination of polarization using social media data, which allows us to unobtrusively observe emergent behaviour \citep{barbera2020use}. This is particularly important for our study, as prior research on the mechanisms of polarization shows that it is a relational, group-level phenomenon \citep{suhay2015explaining,myers1978polarizing}. Our use of social media data allows us to directly observe individuals placing themselves with or against one another in a public forum, affording us the construct validity not available from a survey-based study, which usually rely on aggregated differences in individual responses to infer polarization.

Finally, an additional benefit of our network-based approach is that it is agnostic to language outside of possessing case-specific knowledge, making it readily applicable to different contexts. While natural language processing tools have generally expanded beyond English, there is still a large gap in how they perform across different languages \citep{djatmiko2019review}. Twitter, additionally, is a constrained medium less suitable for textual analysis. Our paper is therefore useful as a guide for a large-scale approach to studying polarization without relying on text.

The remainder of this paper proceeds as follows. In the second section, we describe our theoretical framework and outline a number of expectations. Here, we also discuss the importance of focusing on alignment instead of single-issue polarization. In the third section, we present our empirical approach, which applies network methods to studying public opinion polarization and alignment. We also describe our data and data collection procedure. In the fourth section, we present our findings. We show that, among a set of politically relevant topics, climate politics is not particularly divisive. However, there is strong evidence of partisan sorting, where positions on climate issues are strongly aligned with positions on political parties. With regard to specific issues, climate attitudes have become strongly aligned with attitudes toward immigration. Together, our results indicate that the new universalist-communitarian dimension of political competition emphasized by right-wing populists parties has become an important driver of climate politics polarization, and that electoral cleavages over issues in this political space have become more coherently aligned.

\section{Theoretical Framework and Expectations}
The goal of this is study to look for evidence of polarization in public opinion toward climate politics in the Finnish multiparty system, and to characterize the nature of this polarization. Specifically, informed by the broader literature on political polarization \citep{fiorina2008political,baldassarri2008partisans,layman2006party}, we ask whether the observed climate polarization can be attributed to partisan sorting, and whether this has led to its alignment with other socially salient issues. Our primary focus, therefore, is on how polarization of climate politics aligns with other sources of societal cleavage.

Our emphasis on alignment of cleavages instead of single-issue polarization is motivated by the logic outlined in \citet{baldassarri2008partisans}. To the extent that political issues afford different positions, a populace that is divided over multiple issues is not necessarily harmful to social stability. In open societies, multiple non-overlapping societal cleavages create cross pressures that balance each other out, preventing a single source of conflict from becoming entrenched \citep{coser1956functions}. Highly polarized positions on climate change is not necessarily a problem as long as they do not correlate with opinions on other divisive issues such as immigration. If multiple divisive issues are independent, any two individuals will in expectation agree on half the issues and disagree on the other half, thereby reducing the kind of polarized animosity that makes it difficult to reach agreements. Indeed, while polarization constrains climate communication \citep{zhou2016boomerangs}, studies show that even climate change deniers can be convinced of accepting greenhouse gas mitigation policies when proponents use arguments based on common ground outside of climate politics itself \citep{dryzek2015reason}. However, when these cleavages start to align, cross pressures are reduced, resulting in, at the extreme, two highly antagonized groups with no source of agreement.

Studying what drives public opinion to align across topics (i.e. political parties and issues) is therefore important to understanding polarization. Prior research on this question identified two primary types of alignment \citep{fiorina2008political,layman2006party}. First, partisan sorting refers to the process by which the public sorts itself into entrenched partisan identities, resulting in a populace that is divided along party lines despite holding relatively nonaligned positions across different issues. Second, issue alignment refers to the phenomenon that the electorate's actual preference over policy shifts until ideological divisions become aligned \citep{baldassarri2008partisans}. These two types of alignment can arise independently, but it is within expectations that they reinforce each other under certain conditions \citep{layman2002party}.

It is well established that partisan sorting exists in the U.S. context \citep{fiorina2008political}, as evidenced by the overwhelming alignment of issue positions with partisanship. In certain scenarios, such as among partisans, we observe an associated alignment among issues, but there is no clear evidence of general issue alignment \citep{baldassarri2008partisans}. More specifically, the research shows that for partisan sorting to yield issue alignment, individuals need to possess enough political interest and sophistication, so it usually manifests only among partisans \citep{baldassarri2008partisans,layman2002party}. There are considerably fewer studies of partisan sorting in multiparty systems \citep[e.g.][]{adams2012causes,kevins2018growing}, and here the evidence is highly mixed, depending largely on the case selection and research design specifications.

In the context of climate politics, we expect partisan sorting because climate politics started its life in the public mind as a `niche' issue where expectations for partisan sorting is strongest \citep{adams2012partisan}. As a niche issue, generally understood as specific or narrow topics not directly related to economic policies, climate change was championed, and on the other side antagonized, by nontraditional parties. The Green and to some extent New Left parties have been the primary champions of stronger climate policies. This demand has more recently become mainstream so that established parties have taken stronger positions. This mainstreaming, driven by the mounting scientific evidence on the need to address climate change, as well as the global political process around the United Nations Framework Convention on Climate Change (UNFCCC) has recently been met with rising opposition from populist right parties in many European multiparty democracies. The 2019 parliamentary elections in Finland were in fact coined by some as the `climate elections' \citep{hassinen2019campaign}. This trend of political elites polarizing around the climate issue, coupled with the issue's increasing salience in the public debate, leads us to expect public opinion on climate politics to exhibit partisan polarization tendencies. On the other hand, it is less likely for public opinion on climate politics to align more broadly with public opinion across multiple other issues, simply due to the fact that this phenomenon is generally rare even in highly polarized political environments like the U.S. Indeed, a recent examination of the European Social Survey found that individuals were relatively evenly distributed in terms of support across the two dimensions of climate and social welfare policy \citep{otto2020eco}. From this, we have the \textit{partisan sorting hypothesis}:

\begin{quote}
    \textbf{Hypothesis 1: } Public opinion on climate politics will align more strongly with public opinion on political parties than with other political issues.
\end{quote}

Compared to two-party systems, multiparty systems afford political parties the flexibility to be more eclectic in their chosen issues \citep{meyer2013mainstream}. This is especially so for niche parties, which tend to emphasize issue ownership instead of general competence in governing. Here, \citet{adams2012partisan} find that niche parties across 14 European countries and their supporters tend to exhibit more partisan sorting than mainstream parties. Further, and especially relevant to our focus on climate politics, research on European politics demonstrates the rise of a cultural dimension to political conflict in addition to the traditional left-right distributive divide \citep{kriesi2006globalization}. This new dimension has been characterized as a universalist-communitarian divide focused on the role of the community as an organizing principle of society, with the communitarian populist right arguing against the global inclinations of the universalist New Left \citep{bornschier2010new}. 

On the climate issue, right-wing populist parties in Finland and elsewhere in Europe have positioned themselves as main opponents in the debate \citep{lonnqvist2020polarization,lockwood2018right,gemenis2012politics}. They largely take the communitarian stance that their countries should not be responsible for bearing the costs of climate change mitigation, in many cases labeling the climate movement, with associated international processes around the UNFCCC, as an universalist movement that encroaches upon national sovereignty and sufficiency \citep{forchtner2018being,forchtner2015nature}. Given the increasingly salient role the populist-right has taken as climate policy opponents, we have the \textit{universalist-communitarian hypothesis}:

\begin{quote}
    \textbf{Hypothesis 2: } Partisan sorting of climate politics will be driven primarily by populist parties and their opponents.
\end{quote}

Finally, we return to consider the relationship between partisan sorting and issue alignment in light of the preceding discussion on right-wing populist parties. As noted above, prior research indicates that issue alignment can be a product of partisan sorting given suitable conditions \citep{baldassarri2008partisans,layman2002party}. Specifically, individuals with sufficient political interest or sophistication to overcome the difficulties of identifying their parties' positions across multiple different issues will adjust their own ideological positions on these issues accordingly, thereby leading to issue alignment.

Here, we point to immigration as a potential topic for climate politics to become aligned with. Even more than climate politics, immigration skepticism is an issue heavily championed by right-wing populist parties in Europe \citep{rydgren2008immigration,betz1993new}. This issue fits directly into the universalist-communitarian debate, and is a central mandate for many of these parties \citep{rydgren2008immigration}. From the perspective of political parties, especially niche parties, linkage across issues can be advantageous as it prevents them from being seen as single-issue parties \citep{mudde1999single}, and can consolidate their support base against any particular issue from becoming irrelevant. There is evidence of this behaviour in Finland, where a recent study found that political elites from both the right-wing populist Finns Party and the Green League have become increasingly aligned on their respective positions toward climate and immigration politics \citep{lonnqvist2020polarization}. Given this move, and the high salience in the links between these two issues via the new universalist-communitarian debate, we expect that public opinion on climate and immigration politics will be aligned with each other in addition to exhibiting partisan sorting. This leads to the \textit{immigration alignment hypothesis}:

\begin{quote}
    \textbf{Hypothesis 3: } Public opinion on climate politics will align with public opinion on immigration, but not with traditional left-right distributive politics.
\end{quote}

To recapitulate, our expectations are that polarization in climate politics exhibits strong tendencies of partisan sorting, which leads to public opinion on climate politics aligning strongly with attitudes toward right-wing populist parties and their opponents, and toward the immigration issue contended by these parties. To examine public opinion on these topics (i.e. parties and issues), we focus on public displays of agreement from the Twitter platform in the months surrounding the 2019 Finnish parliamentary and European Parliament elections. Before moving to presenting our empirical approach, we provide a brief background on our case and contextualize our expectations in terms of climate politics and political polarization in Finland and Finnish Twittersphere in particular.

\subsection{Climate Politics in Finland and Finnish Twittersphere}
The Finnish political system has traditionally been based on a tripartite corporatist collaboration between employers, employees, and the government. This has made politics more consensual compared to the two-party system of the U.S. \citep{arter2015scandinavian}. Finnish governments often consist of multiple parties from both the left and the right. This tends to lessen inter-party conflicts because parties expect to enter government with one another at some point in the future. However, the recent rise of a right-wing populist party, the Finns Party, along with the falling popularity rates of traditional parties, has introduced new cleavages. The Finns Party has become almost as popular as the traditional mainstream parties, the center-left Social Democrats and the center-right National Coalition, making them less of a niche party at least in recent years.

Despite the history of consensus politics in Finland, there is evidence of polarization and observable levels of alignment between issues in specific contexts \citep{lonnqvist2020polarization}. In particular, prior research showed that there are conflicts over climate change mitigation policies in Finland. Business actors, trade unions, and some governmental organisations think that economic growth is more important than climate change mitigation and this stance has met resistance from environmental, civil society actors and political parties from the left of the spectrum \citep{teravainen2010political, gronow2019cooptation}. These findings, however, remain generally limited to political and economic elites.

On April 14 and May 26, 2019 respectively, Finland held its parliamentary election and European Parliament election. During the lead up to these elections, climate politics was among the salient debates, thereby providing us with a suitable opportunity to directly observe the formation, polarization, and alignment of public opinion on the Twitter platform. 

Twitter, of course, is not representative of the general population. While there are clear advantages of studying public opinion through Twitter \citep{barbera2020use}, it remains imperative to explicitly contextualize our expected findings in terms of what we understand about Finnish Twittersphere. First, note that use of social media in Finland is high \citep{statisticsfinland2019}. However, there is a clear skew toward the younger population (over 84\% for those under 45, 11\% over 74, and 61\% overall), with the Twitter user group being particularly overrepresented by more politically inclined and urban individuals \citep{vainikka2015tviittien}. Finally, \citet{nelimarkka2020platformed} found that opinion-sharing and position-taking were prominent features of Finnish political Twittersphere in the lead up to the 2015 Finnish parliamentary elections, which bodes well for our approach of measuring public opinion using the platform.

Given the preceding discussion, our findings about how climate politics exhibit tendencies of partisan sorting and issue alignment should be understood as applying to a more politically sophisticated and active segment of the population. As discussed above, under such conditions we should be more likely to observe issue alignment \citep{baldassarri2008partisans}. At the same time, it merits noting that a U.S.-based study found that younger individuals with more education (who are overrepresented on Twitter) tend to be less likely to align with their party on the climate issue \citep{ross2019polarization}, which might decrease observable partisan sorting and issue alignment.

\section{Empirical Approach}
In this study, we take a network approach to studying public opinion. Specifically, we measure individuals' political attitudes by looking at the networks of agreement among individuals communicating with one another about given topics. Using publicly available data from the Twitter platform, we construct networks where individuals agreeing with each other are connected, leading to the emergence of groups with the same stance on topics. The resulting networks allow us to apply global network measures to study polarization and cross-topic alignment, which we describe in more detail in section \ref{sec:measures}. This logic of aggregating relational behaviour on social media to infer positions has been applied elsewhere to measure political ideology \citep{barbera2015birds,conover2011predicting}, and has a number of advantages. 

First, data from Twitter is publicly accessible directly through its application programming interface (API). We can therefore collect behavioural data without interacting with our subject pool. Beyond the ethical advantages of this approach, we are also able to minimize all sources of interviewer effects associated with surveys, including measurement errors as well as unit and item nonresponse \citep{west2017explaining}. As noted above, inferences about polarization is particularly susceptible to measurement error introduced by respondents' tendency to choose the middle category \citep{hetherington2009putting,moors2008exploring}. At the same time, minimizing nonresponse means that we are able to obtain a near census of the system (with the exception of private accounts). 

Related to this, the second advantage is that we are able to observe the system in its entirety over a continuous temporal period, thereby capturing outcomes of emergent processes and other behaviours that are constrained in individual-level surveys. This is particularly important for our study, as polarization and alignment across topics are system-level phenomena driven by relational, group behaviour \citep{suhay2015explaining,myers1978polarizing}. Whereas survey-based studies infer the level of polarization using aggregated differences in individual responses, we are able to directly observe individuals placing themselves with or against one another in a public forum across multiple topics.

\subsection{Data Collection}
We collected tweet and retweet data over the entire election and post-election period, from March 1 to July 31, 2019. Data collection was done via the Twitter API which returns all tweets satisfying user-specified conditions \citep{twitterdev2020}. Specifically, we filtered the Twitter stream by a set of 317 hashtags, meaning that all tweets and retweets which include at least one of these hashtags entered our data set, subject to a small number lost to random connection drops. These hashtags were selected based on relevance to the parliamentary and European Parliament elections in Finland. 

Notably, we identified eleven topics to focus our examination on. Six of these are political parties and five of them are socially salient issues, including climate politics. These topics are outlined in \autoref{tab:topics}. Each of these topics has its own set of associated hashtags that is a subset of the overall 317. These topic-specific subsets, which we use to measure public opinion on the topics, are mutually exclusive from each other. The hashtags used for these eleven topics are translated into English and presented in \autoref{appendix:keywords}. 

\begin{table}[!htb]\renewcommand\arraystretch{1.2}
    \centering \footnotesize
    \begin{tabular}{l l | l  l} \hline \hline
        \multicolumn{2}{c}{\textbf{Issues}} & \multicolumn{2}{c}{\textbf{Parties}}  \\ \hline
        Climate & Immigration & Social Democrats (SDP) & Finns  \\
        Social Security & Economic Policy & National Coalition  & Green League \\
        Education & & Centre & Left Alliance \\
        \hline
    \end{tabular}
    \caption{Topics Used in this Study}
    \label{tab:topics}
\end{table}

We selected the four socially salient issues in addition to climate politics based on their importance to these elections, but also with an eye toward capturing the traditional left-right divide, the new universalist-communitarian divide, and a major campaign issue. Immigration was particularly salient because the anti-immigrant Finns Party had advanced significantly in the polls. Recent events prior to the election period, particularly police investigation into a series of sex crimes by immigrant men, had created public controversy around the issue. Together with climate politics, they constitute the majority of the current universalist-communitarian debate in Finland. Economic policy and social security are issues that traditionally divide the left and the right, and tends to be strongly present in campaign debates. Social security was particularly salient in this electoral period due to the sitting-government's failed reform efforts. Education became a major campaign issue during these elections due to the cuts in funding by the previous government and strong demands from the part of student organizations and other actors that the cuts be reversed under the next government. 

\subsection{Network Construction}
We divided the data collected into three periods, pre-election (March 1 to April 14), inter-election (April 15 to May 26), and post-election (May 27 to July 31). Using data from each period, we constructed a set of endorsement networks for our eleven topics. Following prior work \citep[e.g.][]{garimella2018quantifying,barbera2015tweeting}, we constructed these endorsement network using retweets, which are unmodified sharing of original tweets to the retweeting user's timeline (i.e. without accompanying text). While some account have bios indicating that their retweets are not endorsements, prior research showed that the majority of users retweet messages they agree with, find trustworthy, or endorse \citep{metaxas2015retweets}, and that they tend to retweet others with the same political leaning \citep{barbera2015tweeting}.

Each resulting network has a node set containing all users who posted an original tweet containing at least one hashtag related to the given topic and all users who retweeted at least one of these tweets. Undirected ties on the network indicate the linked nodes have at least one instance of retweet between them on the given topic. Descriptive statistics for these networks are presented in \autoref{tab:networks}.

\begin{table}[!htb]\renewcommand\arraystretch{1.2}
    \centering \footnotesize
    \begin{tabular}{l r r@{\hskip 2em} r r@{\hskip 2em} r r} \hline \hline
        & \multicolumn{2}{c@{\hskip 2em}}{\textbf{Pre}} & \multicolumn{2}{c@{\hskip 2em}}{\textbf{Inter}} & \multicolumn{2}{c}{\textbf{Post}} \\
         \textbf{Topic} & \multicolumn{1}{c}{$N_{v}$} & \multicolumn{1}{c@{\hskip 2em}}{$N_{e}$} & \multicolumn{1}{c}{$N_{v}$} & \multicolumn{1}{c@{\hskip 2em}}{$N_{e}$} & \multicolumn{1}{c}{$N_{v}$} & \multicolumn{1}{c}{$N_{e}$} \\ \hline
         Climate & 14510 & 42329 & 6445 & 13624 & 6058 & 11775 \\ 
         Immigration & 2839 & 6878 & 1701 & 3463 & 2367 & 4968 \\ 
         Social Security & 7276 & 16748 & 3486 & 5815 & 3291 & 5446 \\ 
         Economic Policy & 2936 & 4420 & 2348 & 3316 & 2461 & 3638 \\
         Education & 6992 & 12199 & 3690 & 6125 & 3737 & 5864 \\ 
         Social Democrats & 2004  & 4457 & 1256 & 2719 & 601 & 961\\ 
         Finns & 1515 & 4187 & 1497 & 2971 & 1645 & 2850\\
         National Coalition & 2557 & 6458 & 1357 & 2624 & 806 & 1297\\ 
         Green League & 2118 & 4402 & 1439 & 2565 & 1657 & 3660\\
         Centre & 1489 & 2740 & 1324 & 2068 & 933 & 1609\\
         Left Alliance & 1120 & 2341 & 704 & 1192 & 453 & 609 \\ \hline
         \multicolumn{7}{l}{\begin{minipage}[t]{.6\textwidth}
         $N_{v}$ refers to number of nodes; $N_{e}$ refers to number of ties. \end{minipage}}
    \end{tabular}
    \caption{Descriptive statistics for endorsement networks by topic and period.}
    \label{tab:networks}
\end{table}

\subsection{Measures} \label{sec:measures}
We approach the measurement question by considering, if a political system is marked by polarized individuals, what would the observable manifestations be? Drawing on studies from computational social science \citep[e.g.][]{garimella2018quantifying,bright2018explaining,barbera2015tweeting}, we look for evidence of polarization in political communication patterns at the systemic level, using global network measures (i.e. network statistics computed over the entire network) of group fragmentation and cross-network group alignment. We are not the first to measure polarization using differential densities of inter- and within-group agreement, but our extension to studying alignment using similarity measures of group partitioning presents a theoretically motivated bridge between the numerous studies of within-issue polarization and the broader political polarization literature.

\subsubsection{Within-topic Polarization} \label{sec:polarization}
We measure system-level trends in public opinion polarization by considering patterns of agreement between opposing groups. Specifically, we measure polarization as the relative density of in-group agreement to out-group agreement. We assess these patterns on our endorsement networks, which are a subtype of general communication networks where all ties are publicly conveyed indications of agreement. Practically, this means subsetting all interactions on Twitter to only retweets, leaving out other commonly-identifiable ties such as follows, replies, mentions, and quotes.

We begin by partitioning the network with the objective of obtaining two similarly-sized groups that have the least amount of ties between them. In the context of endorsement networks, this captures, as we intend, two groups of individuals with the lowest level of agreement between them. Restrictions on the balance between group size is required because the unrestricted algorithm will lead to a trivial partition with a single degree-one node in one group and the rest of the nodes in the other group.
While this satisfies the objective of minimizing inter-group ties to exactly one tie, it clearly does not capture our intentions. On the other hand, forcing the partitions to be perfectly balanced when real groups are not the same size will result in the partition to divide the larger group, which incorrectly inflates inter-group agreement relative to within-group agreement. We therefore set a 3:7 maximum imbalance constraint on the the partitioning algorithm \citep[METIS, see][]{karypis1998fast}. Community detection for networks is a deep field \citep{fortunato2016community}; for our present problem of finding two groups with the least amount of agreement, the METIS algorithm has a straightforward and intuitive interpretation, and has been shown to work well with retweet networks \citep{garimella2018quantifying}. 

The resulting partition is used to organize the adjacency matrix of the network into a block matrix, where the within-group ties are on the main diagonal blocks and the inter-group ties are on the off-diagonal blocks. Then, given the block matrix $B$, we calculate our polarization score $P$ as 
\begin{equation}
  P = \frac{B_{aa}+B_{bb}-B_{ab}-B_{ba}}{B_{aa}+B_{bb}+B_{ab}+B_{ba}}\,,\label{eq:polarization}
\end{equation}
where $B_{ij}$ is the density of ties in block $ij$. Within each block, the density of ties is defined as the number of observed ties $c$ divided by the total number of possible ties $n$ (i.e. $c/(n_i(n_i-1)/2)$ for within-group blocks and $c/n_in_j$ for inter-group blocks). 

Our score extends the EI-index \citep{krackhardt1988informal}, which is based on the distribution of ties linking nodes within and between groups. The denser the within-group links relative to inter-group links, the more polarized the system. The EI-index has been used elsewhere in studies of group-moderated communication patterns \citep[e.g.][]{bright2018explaining,hargittai2008cross}, but does not handle unequal group sizes well. Specifically, because the EI-index is based on the raw count of ties in the blocks, it increases monotonically with inequality in the groups' sizes; off-diagonal blocks of a block matrix necessarily decrease in size relative to the main diagonal blocks as the latter becomes uneven. This undesirable property holds even for networks that have denser inter-group ties than within-group ties (i.e. those with negative EI-index when the groups are equal). Our formulation accounts for different group sizes by using the density of ties within each block, thereby maintaining a constant score as long as the density remains the same regardless of block size. When the two groups are the same size, $P$ reduces to the EI-index.

\subsubsection{Alignment across Topic Pairs}
We measure alignment across a pair of topics as how similar the two topics are in the way accounts position themselves around each topic, based on the partitioning step described in section \ref{sec:polarization}. Specifically, for a given pair of topic endorsement networks, we take the subset of accounts partaking in discussions on both topics, then compare the extent to which the two pairs of groups overlap. The more each group from the first topic overlaps with one of the groups in the second topic but not with the other, the more similar the group memberships are, and the more we consider these topics to be aligned. 

\begin{figure}[!htb]
    \centering
    \begin{tikzpicture}
    \node (design) at (-3.9, 0) {\includegraphics[width = 0.5\textwidth,trim = {1cm 1cm 1cm 1cm}, clip]{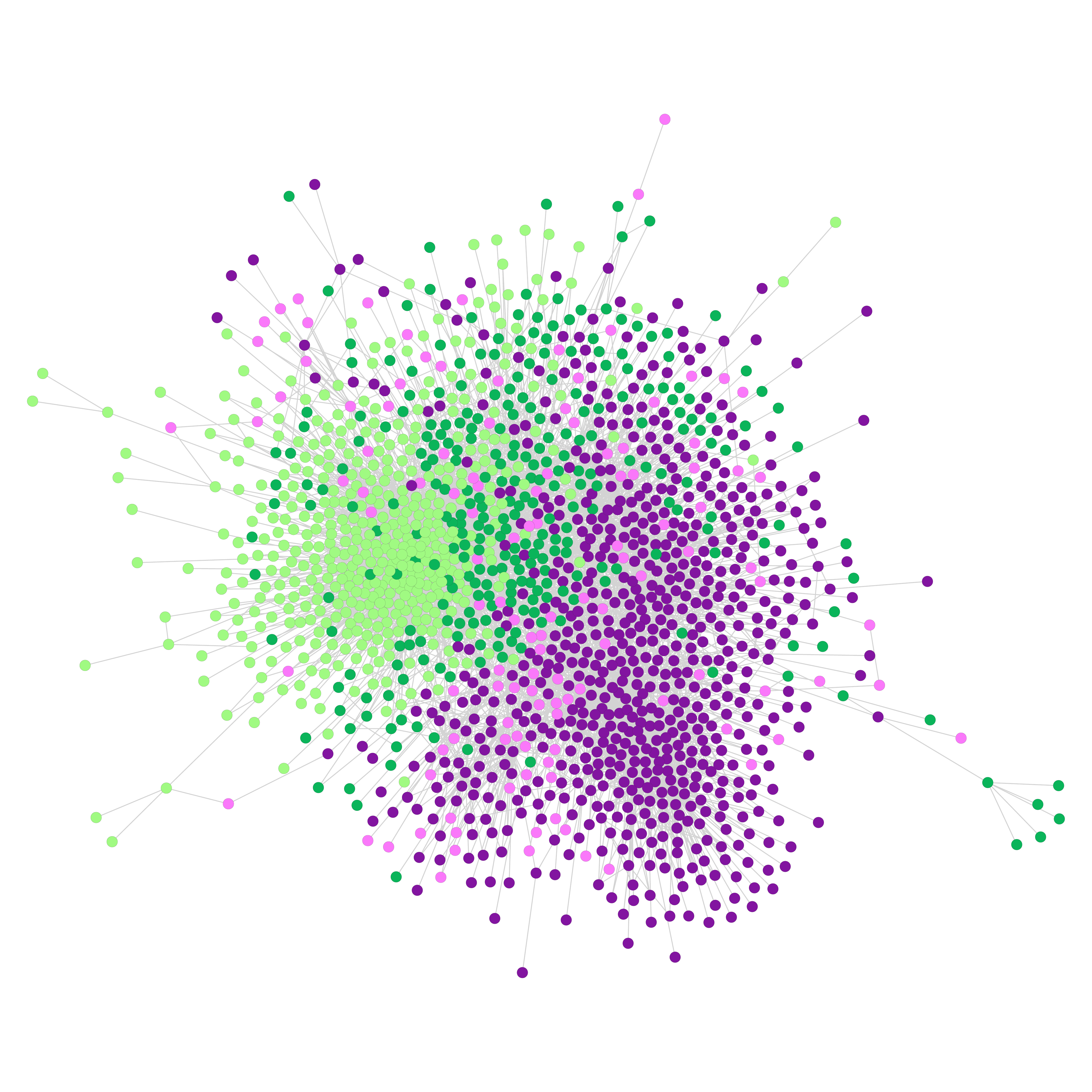}};
    \node (design) at (3.9, 0) {\includegraphics[width = 0.5\textwidth,trim = {1cm 1cm 1cm 1cm}, clip]{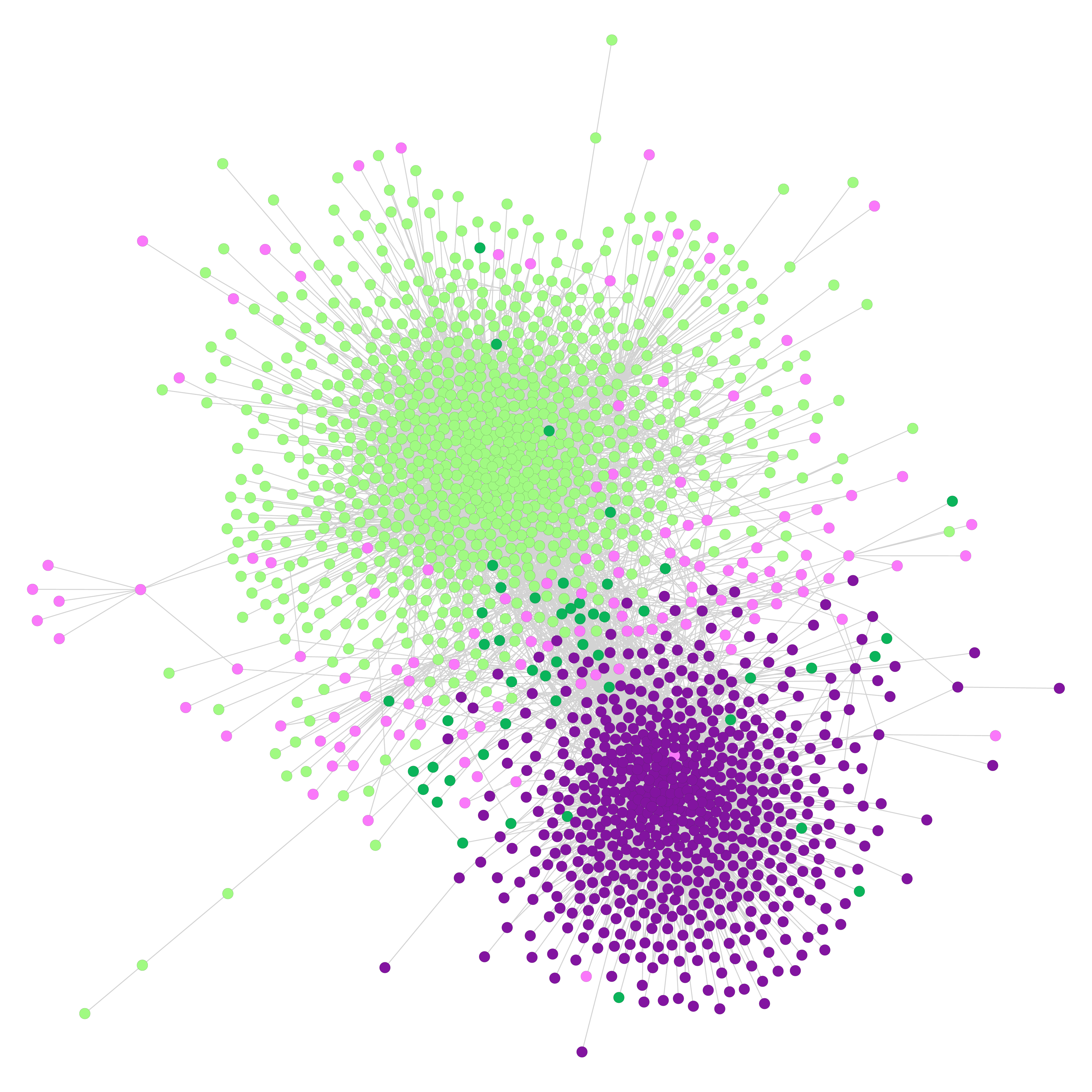}};
    \node (design) at (-6.8, -3.2) {\includegraphics[width = 0.135\textwidth,trim = {0cm 1.5cm 0cm 0cm}, clip]{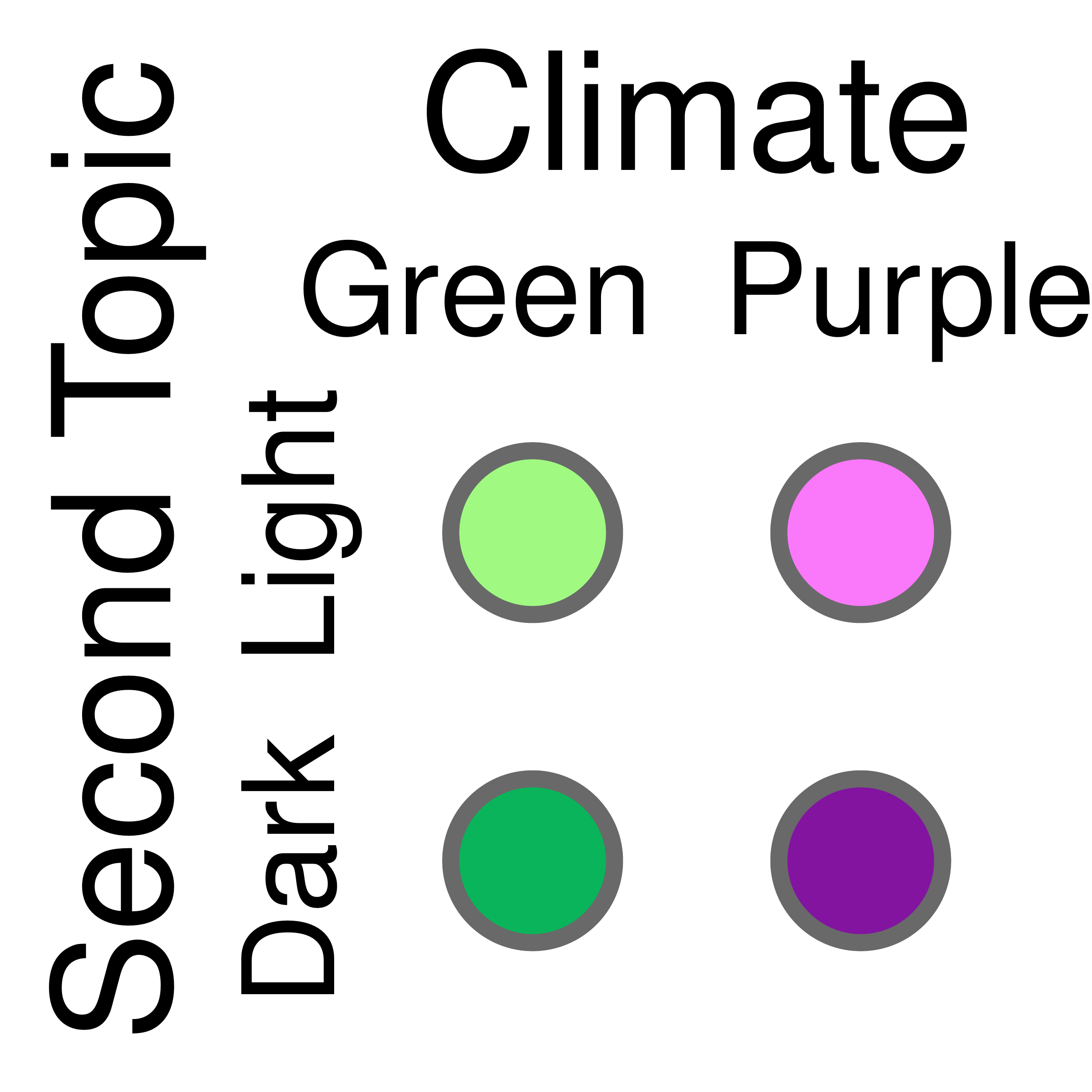}};
    
    \node[label={left, gray}:{\footnotesize \sffamily{Climate-Economic Policy}}] at (-3, 3.7) {};
    \node[label={left, gray}:{\footnotesize \sffamily{Climate-Immigration}}] at (4.2, 3.7) {};
    \end{tikzpicture}
    \caption{Joint endorsement networks with labeled groups. The left-hand side figure is the joint endorsement network of climate politics and economic policy. The right-hand side figure is the joint endorsement network of climate and immigration politics. On both networks, positions on climate are labeled using green and purple (in both shades), while the second topic is coded with dark and light shading of the two colours.}
    \label{fig:overlapping}
\end{figure}

We illustrate this concept with two examples in \autoref{fig:overlapping}. On the left is the joint endorsement network of climate and economic policy, issues that are relatively unaligned. On this network, accounts taking the two different positions toward climate politics are respectively coloured with green (dark and light) and purple (dark and light); on the other hand, the two positions toward economic policy are differentiated using dark (green and purple) and light (green and purple) shades. The two topics, climate politics and economic policy, has relatively low alignment, so the groups are mixed. Accounts in the green climate group are relatively evenly distributed into the light and dark economic policy groups, and those in the purple climate group are also relatively evenly distributed into the two economic policy groups. Similarly, we can see that the dark economic policy group is relatively evenly distributed into the green and purple climate groups, and so on. As discussed previously, where two topics have low alignment, there are more common sources of agreement. For example, a dark green account and a dark purple account disagree on climate politics but agree on economic policy, reducing entrenched identities that prevent cooperation in policy-making. Conversely, on the similarly-labeled joint endorsement network of climate and immigration on the right, alignment is high, and accounts are sorted into clearly defined groups. Most members of the green climate group are members of the light immigration group, and most members of the purple climate group are members of the dark immigration group.

\begin{figure}[!htb]
    \centering
    \includegraphics[width = 0.4\textwidth]{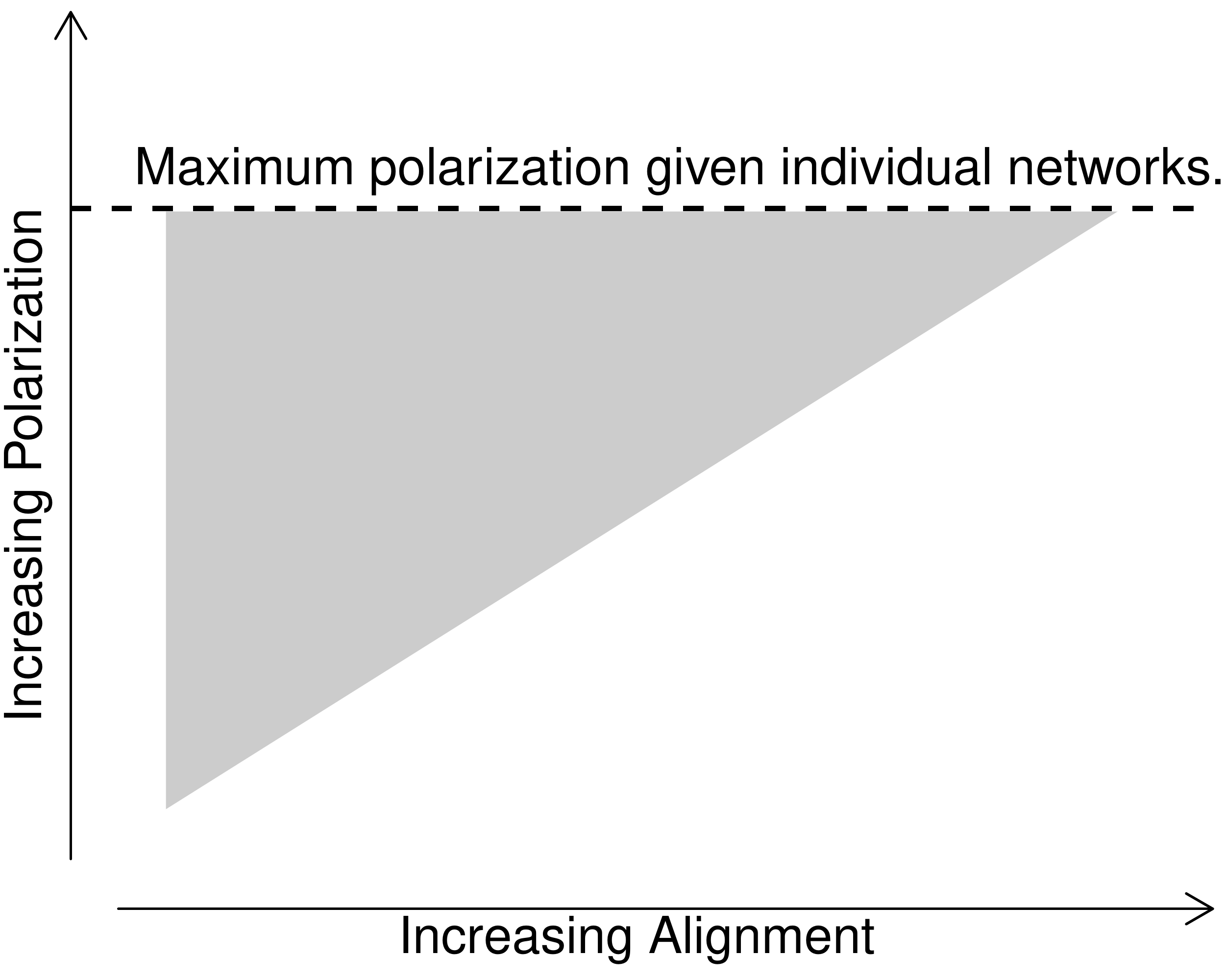}
    \caption{Schematic illustration of the relationship between two networks' alignment and the polarization of their joint network. The horizontal dotted line indicates the maximum polarization possible for the given joint endorsement network.
    }
    \label{fig:align_to_pol}
\end{figure}

Following the discussion outlined in our theoretical framework, a valid measure of topic alignment must have the property that as alignment increases, polarization in the given system (i.e. joint endorsement network) will also increase even when holding within-issue polarization constant \citep{baldassarri2008partisans}. To see why a measure based on overlapping group memberships satisfies this condition, consider the following. Begin with two topic-specific endorsement networks with identical node sets, both with $P > 0$ (i.e. on these networks, the density of ties within communities is higher than across). Next, create a joint-endorsement network by collapsing the original endorsement ties over the common node set. Consider the limiting case where the nodes' group memberships on the two original networks are identical. Because the partitioning step applied to the two original networks already minimized inter-group ties, the number of inter-group ties on the new joint network will remain at the lowest possible value for the given tie set. From here, decreasing the overlap between the two original group memberships will mean that the ties on the joint network will be increasingly evenly distributed across any two groups on the network. The relationship between alignment and polarization in the joint network is therefore as schematically shown in \autoref{fig:align_to_pol}. While it is possible to construct convoluted examples of networks that remain maximally polarized (for the tie set) across all levels of alignment, real-world networks will sit somewhere closer to the lower bound of the shape.

We formally measure similarity in group memberships using the mutual information between partitioning outcomes on different endorsement networks. Mutual information is an information theoretic measure of the similarity between two distributions \citep{strehl2002cluster}. Intuitively, it quantifies how much information knowing the group of an account in one topic gives about the account's group in the other topic. In the context of our study, mutual information captures how certain we can be about, for example, an account's position on climate politics given its stance toward economic policy or immigration. We use a normalized version of mutual information commonly used for comparing network partitions \citep{danon2005comparing,fortunato2016community}, which corrects for unequal group sizes across topics by dividing the raw mutual information by the average of the two partitions' informativeness (i.e. Shannon entropy). This measure is bound between zero and one, where zero means the two partitions are independent and one means they are identical. The climate-economic policy and climate-immigration joint endorsement networks in \autoref{fig:overlapping} have normalized mutual informations of respectively 0.16 and 0.51.

\section{Results and Discussion}
Based on our results, we find evidence to support all three hypotheses outlined in our theoretical framework. Our set of topics are polarized to varying degrees, but there is generally low alignment between them, compared to the extent each of them are aligned with political parties (H1: partisan sorting). The main exception is the strong alignment between climate and immigration politics (H3: immigration alignment). Both of these issues are also strongly aligned with parties that primarily compete in the universalist-communitarian political dimension (H2: universalist-communitarian). Before discussing our alignment results in detail, we first briefly describe the level of within-issue polarization in the system.

\subsection{Within-topic Polarization}
\begin{figure}[!htb]
    \centering
    \includegraphics[width = .9\textwidth]{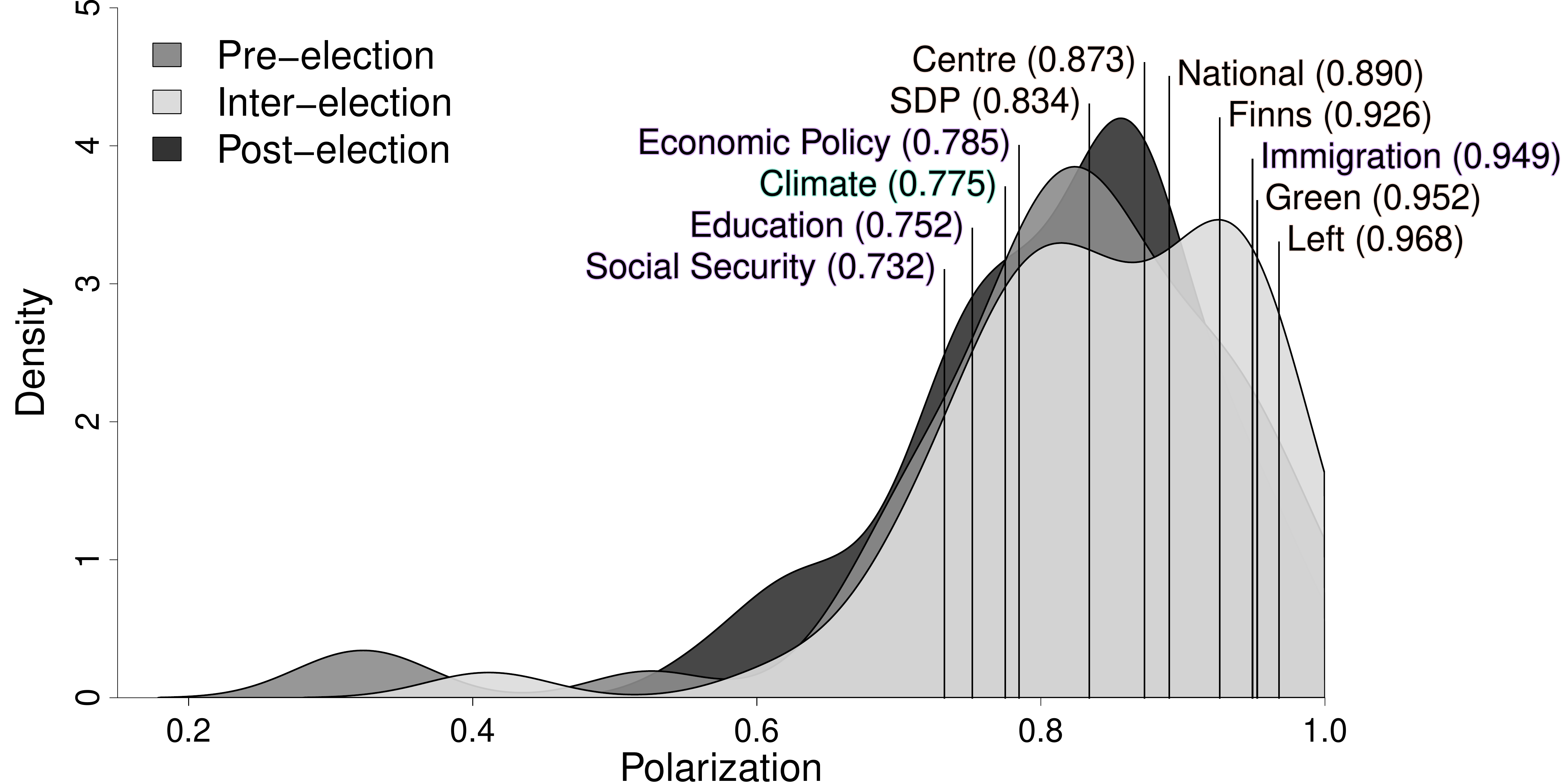}
    \caption{Distribution of polarization by topics. The values for the eleven indicated topics are based on the collection of hashtags outlined in \autoref{appendix:keywords}, averaged across the three periods. The density plots are based on endorsement networks formed by the top fifty hashtags used in each period. These distributions are estimated with kernel density smoothing.}
    \label{fig:ps_dist}
\end{figure}

We report here the polarization of the eleven within-topic endorsement networks. In order to contextualize the polarization scores of these topics, we also estimated the polarization scores for the fifty most common hashtags from our data set in each period, all of which are politically salient topics from the 2019 electoral period. These results are presented in \autoref{fig:ps_dist}. We find that the scores for these endorsement networks center at approximately 0.83. The distributions of $P$ for the hashtag endorsement networks in the three periods appear to be similar, with the largest exception being that the inter-election period is slightly more polarized on the whole. These differences, however, should not be over-interpreted as the most common hashtags are not the same across periods. 

Shown with vertical lines are the polarization scores averaged across periods for each of the eleven topics. Our results indicate that in comparison with the rest of the topics, political parties are among the more polarized. Climate politics is not particularly polarized, ranking above 42 of the hashtag endorsement networks. The other issues are also relatively unpolarized, with the exception of immigration politics, which ranks among the most polarized of the topics. 

\subsection{Topic-pair Alignment}
\begin{figure}[!htb]
    \centering
    \includegraphics[width = 0.7\textwidth]{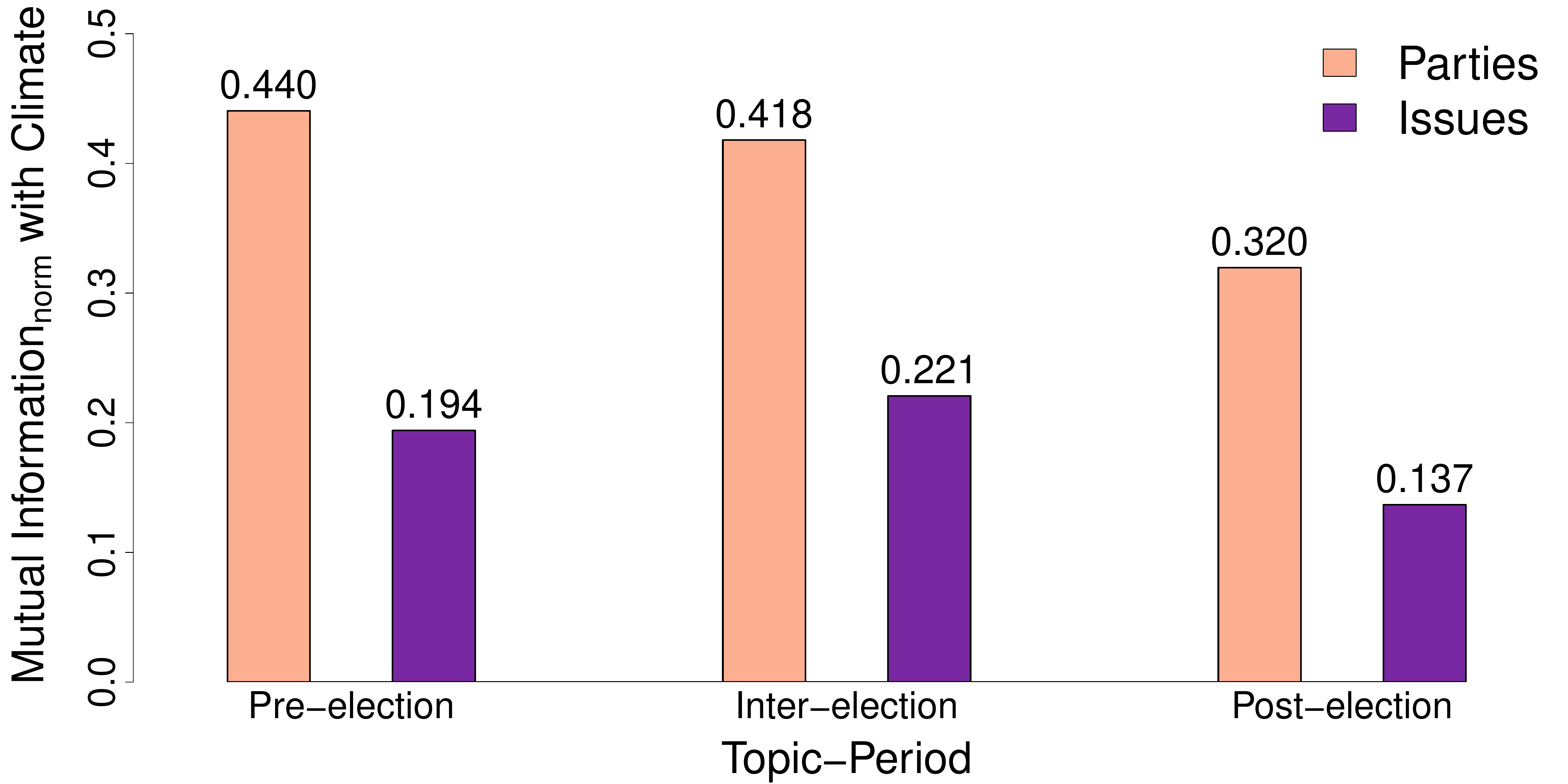}
    \caption{Averaged normalized mutual information for climate politics with party topics and issue topics.}
    \label{fig:alignment_main}
\end{figure}

Next, we focus on the alignment between different topic-pairs. We begin with a broad overview by reporting averaged results for climate politics' alignments with all issue networks and all party networks. Next we move to examine the specific topics in greater detail. According to our \textit{partisan sorting hypothesis}, in our initial examination we expect to find that public opinion on climate politics exhibits tendencies of partisan sorting (i.e. alignment between climate politics and party networks) but not general issue alignment. The results, shown in \autoref{fig:alignment_main}, support this hypothesis. 

First, in all three periods, the climate endorsement network aligns more strongly with the party endorsement networks than with the issue endorsement networks. Second, while we did not explicitly hypothesize about temporal trends (i.e. arising from the electoral cycle), the results are telling. In the post-election period, we find alignment between climate politics and issues in general to be very low. Compared to this baseline, the slightly higher issue alignment during the elections (i.e. pre- and inter-election periods) suggests that the context of national and European Union elections in Finland is enough to raise awareness and interest in political issues such that segments of the public are able to identify and adjust their issue positions to that of their parties', at least in the short term. Together, these results also indicate that despite the overrepresentation of politically active and sophisticated segments of the population in the Finnish Twittersphere, general issue alignment is unlikely outside of specific politically-charged contexts.

\begin{figure}[!htb]
    \centering
    \includegraphics[width = 1\textwidth]{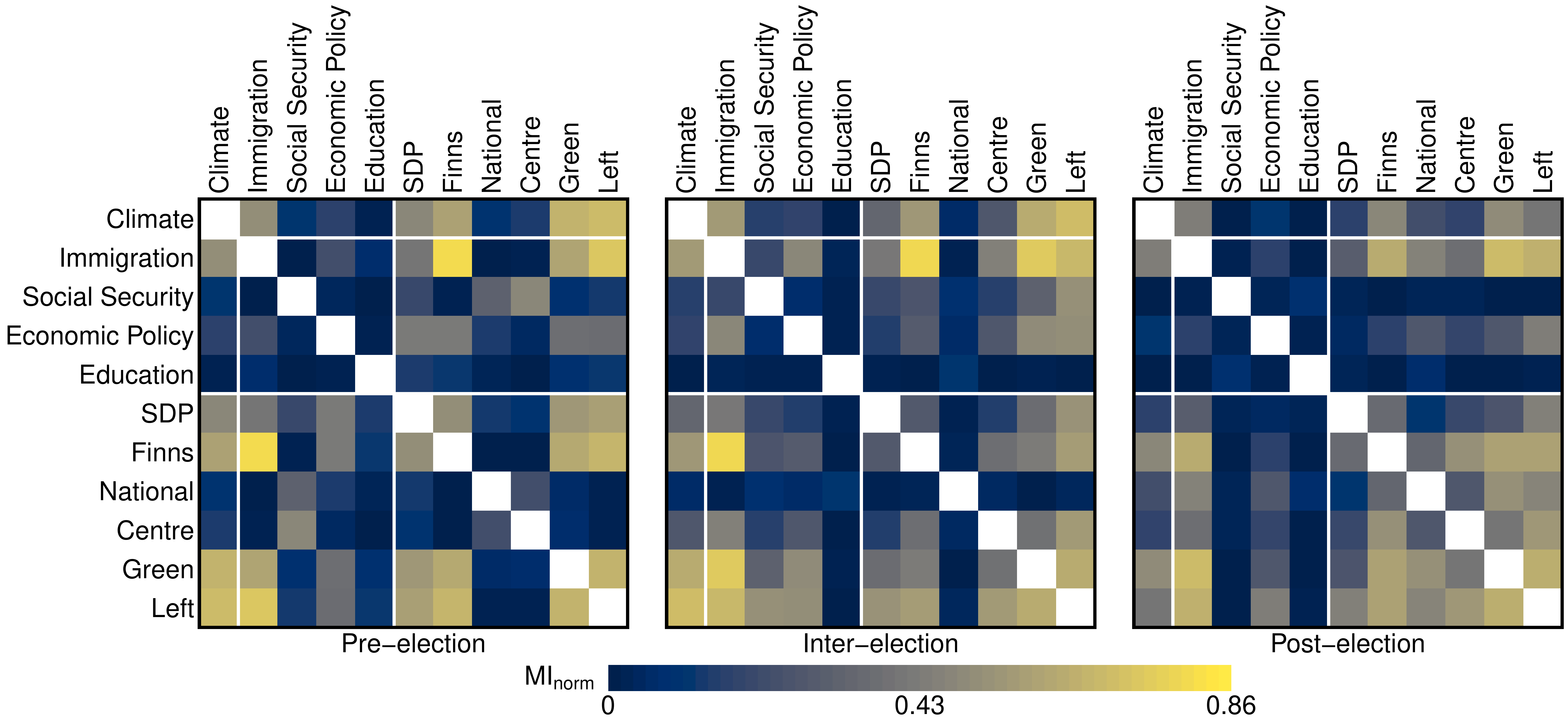}
    \caption{Normalized mutual information for all topic-pairs in each of the three periods. Scale runs from 0 to 0.86. A network's alignment with itself is always 1, and therefore omitted from the figure.}
    \label{fig:mi_map}
\end{figure}

Next, we move beyond the averaged alignments and report the relationships between our specific topics. These results are presented in \autoref{fig:mi_map}, which shows for each period the alignment between every topic-pair. Alignment between the climate endorsement network and all other topic endorsement networks are the cells in the first row. Here, we find support for both the \textit{universalist-communitarian hypothesis} and the \textit{immigration alignment hypothesis}. 

In terms of partisan sorting, while public opinion on climate politics aligned with positions on parties in general, there is considerable variation when parties are considered separately. Specifically, positions on the Finns, the Green League, and the Left Alliance (and to a lesser extent the Social Democrats) appear to be what drives partisan sorting in climate politics. On the other hand, the National Coalition and the Centre Party appear to have little to do with climate politics. These patterns are most apparent in the pre-election period, but they hold across all periods, despite, again, alignment levels generally decreasing following the elections. As we expected, public opinion on immigration also aligned strongly with climate politics and these political parties, indicating a clustering of alignments between topics that operate primarily within the universalist-communitarian dimension.

There are two exceptions to note here. First, positions on the Social Democrats also aligned with climate politics during the elections. In fact, in the pre-election period, we observe partisan sorting specific to the Social Democrats across most topics, if only moderately. During the campaign, the Social Democrats took, for the first time, a strong pro-climate position in response to pressure from the media, the climate movement, and other parties in the wake of the release of the IPCC's influential 1.5 degree report \citep{ipcc2018global}. This is the likely explanation for the alignment of positions on the Social Democrats and climate. The overall centrality of the Social Democrats, in turn, can be explained by the fact that they led the polls throughout the campaign and were, consequently, challenged by other parties on a wide range of issues. Second, economic policy, which is traditionally a left-right debate, aligned more closely with this cluster of topics. We propose this is due to the Finns Party's attempt, similar to other populist-right parties in Europe, to make the discussion surrounding government debt a part of its issue portfolio. Even accounting for these exceptions, clustering along the universalist-communitarian dimension appears to be strong.

The temporal trends from the earlier examination also manifest here. In the post-electoral period, we find that issue alignment drops to almost none, with the exception of alignment between climate and immigration which remains consistently strong. Notably, public opinion toward all parties becomes relatively aligned with one another. This suggests that the post-election government formation led to parties shifting into a center-left bloc that formed the government (comprising the Social Democrats, Centre, Greens, Left and the Swedish People's Party), leaving the center-right National Coalition and populist-right Finns Party in opposition. Partisan competition appears to become unidimensional, and climate and immigration politics are the only issues that remain sorted along partisan lines.

\subsection{Topic-clustering along Political Dimensions}
\begin{figure}[!htb]
    \centering
    \includegraphics[width = 0.7\textwidth]{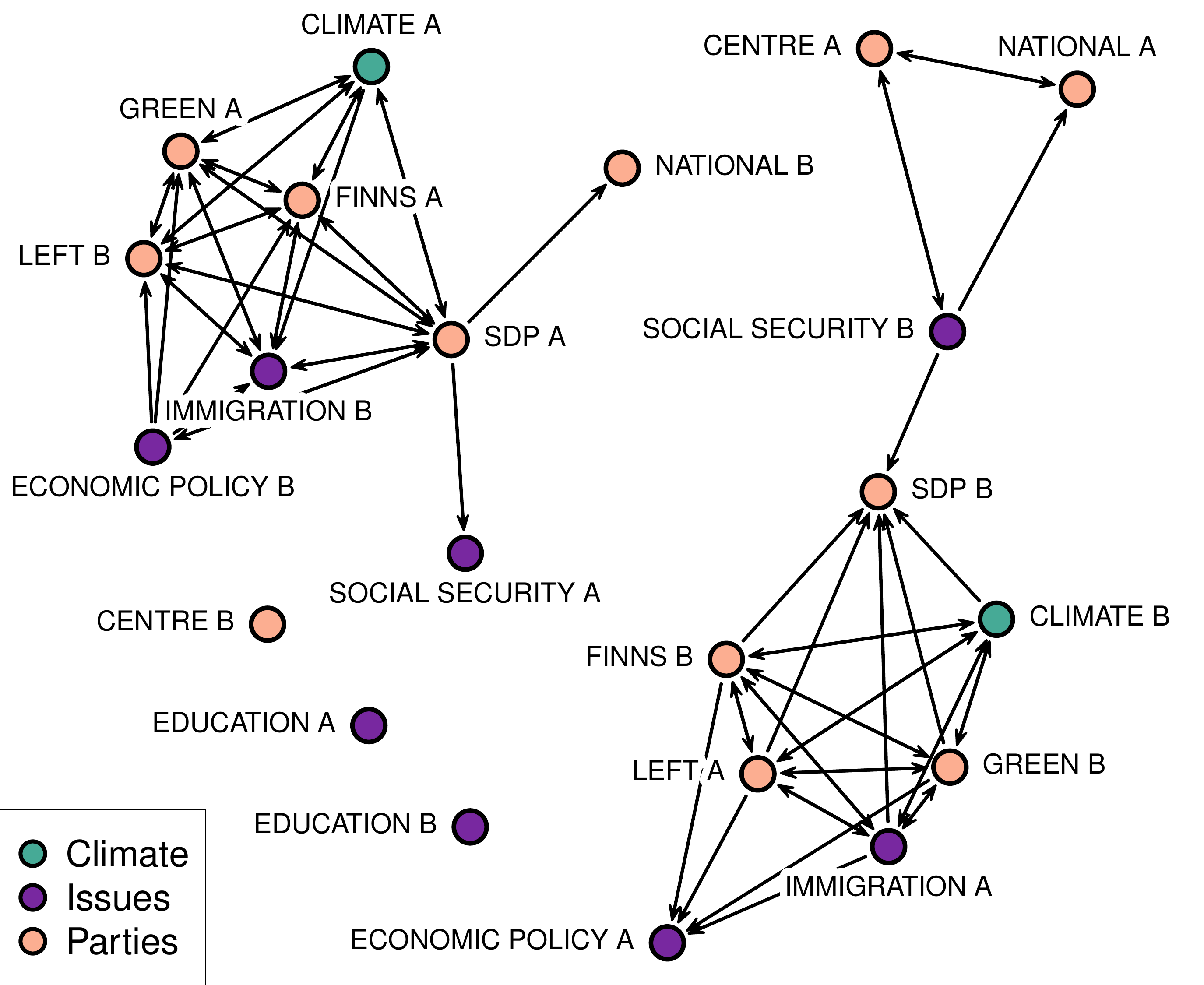}
    \caption{Visualization of clustering in topic alignments. Shown are the alignment patterns among topic-groups in the pre-election period. Nodes represent the topic-group described in the main text. A directed tie indicates that membership in the sender topic-group strongly predicts (greater than 86\%) membership in the receiver topic-group. 
    }
    \label{fig:overall_alignment}
\end{figure}

Beyond our hypotheses, our results allow us to explore the relationship between partisan sorting and issue alignment in multiparty systems more generally. Here, we note that political topics in Finland appear to be subject to a kind of sorting along specific political dimensions (i.e. universalist-communitarian versus left-right), making the overall level of alignment in the system higher than simple partisan sorting but lower than broad general issue alignment.

To illustrate this more clearly, we use the overlap-as-information concept that mutual information is based on, and construct the network plot in \autoref{fig:overall_alignment}. In this figure, we show explicitly the relationship among different groups across all topics in the pre-election period. Nodes in this network are groups from the eleven different topic endorsement networks. For example, the two groups in the climate endorsement network (in green) are labeled as `Climate A' and `Climate B'. A directed tie in this network indicates that an account belonging to the sender group means it is likely to be in the receiver group. Specifically the probability of being in group$_j$ for the receiver topic given membership in group$_i$ for the sender topic, conditional on the size of group$_j$, is above 86\%, a threshold manually selected to most clearly show meaningful clusters among the topics. More generally, we can understand two connected groups having high overlap in membership, and therefore are closely related. 

The immediately observable feature of this network is that, with the exception of three isolated nodes, the groups are clustered into two components, with one group from each topic on each side. This is indicative of a higher level of topic alignment beyond just pairwise alignment. Closer inspection reveals that these components are internally organized by the two dimensions governing European politics. First, as noted above, the Social Democrats, due to their front-runner status throughout the elections, are tied to topics operating in both political dimensions. In both components, the Social Democrats are connected to a dense cluster of topics, including climate politics, that are part of the new universalist-communitarian dimension. On the other hand, the Social Democrats are also linked to topics from the traditional left-right dimension. This is more clearly observable from the component on the right, where the Social Democrats, along with the Centre Party and the National Coalition, are aligned with the traditional left-right issue of social security.

The results here then provides initial evidence that polarization and alignment in multiparty systems operate in a different manner than in two-party systems. Different from a two-party system where issue alignment driven by partisan sorting can occur only in a single dimension, a multiparty system affords a certain flexibility whereby even if persistent partisan sorting leads to alignment between issues, these outcomes can be localized to specific political dimensions. Specific to our examination, we show that while climate politics exhibited the troubling tendencies of partisan sorting and subsequent alignment with immigration politics, it did not align with positions on social security despite the latter also being an important political issue throughout the electoral period.

\section{Conclusion}
Prior research found evidence that public opinion on climate politics sorts along partisan lines. However, they leave open the question of whether climate politics and other politically salient issues exhibits tendencies for issue alignment, which the general political polarization literature identifies as among the most deleterious aspects of polarization. Using a network approach and social media data from Twitter, we measured public opinion on climate politics and ten other topics during the 2019 Finnish elections as the emergence of opposing groups, overcoming previously identified shortcomings associated with survey-based research on political polarization. 

We found that climate politics is not particularly divisive compared to a set of politically relevant topics. However, there is evidence that it is subject to partisan sorting, where public opinion toward climate politics is strongly aligned with positions on parties, specifically the populist-right Finns Party, the Green Party, the Left Alliance, and (more moderately) the center-left Social Democrats. We also found evidence of issue alignment between climate and immigration politics, which persists throughout the electoral period and after. Together, these findings indicate that the relatively new universalist-communitarian dimension of political competition in Europe has become an important component of the climate debate in Finland.

Our study affords a number of outlooks on the polarization of climate politics. First, while alignment can be highly dynamic and subject to the political cycle, with most issues becoming unaligned with each other after the elections, climate politics remain consistently aligned with immigration politics despite shifting political contexts throughout the electoral period and after, suggesting that this troubling phenomenon is likely to persist. At the same time, there is evidence that the Finnish multiparty system is plural enough such that even where partisan sorting led to issue alignment, it largely occurs in localized contexts; populist parties champion immigration skepticism, and successfully linked the climate issue to their portfolio, but the traditional distributive debate on the left-right dimension remains separate from climate politics. Importantly, this means that politically salient issues in this other dimension such as social security spending are therefore unlikely to be aligned with climate politics, keeping the overall polarization in the system to a manageable level.

Finally, our study should be read as an intended contribution to the broader literature on climate politics polarization. In this regard, while our focus on the Twitter platform affords us with internally valid measures of polarization and alignment in the system, the extent to which findings about this particular system can be generalized to the broader system it is nested in remains an open question. In addition to previously discussed differences between the Twitter and general populations, group behaviour across different public forums including social media platforms can vary by features of the contexts, so we might not observe the same phenomenon manifesting in, for example, Facebook discussions. In this sense, we are not able to directly compare our results to existing studies which are more uniform in their methodology. Still, we believe our study demonstrates first the importance for future work to consider issue alignment more deeply when studying climate polarization, and second the advantages of doing so using our proposed approach. We look forward to more general conclusions and greater understanding of climate politics polarization as the body of work in this area continues to grow.

\clearpage
\newpage
{\small \singlespacing
\bibliography{polarization}
}
\newpage
\appendix
\section{Data Collection Keywords}\label{appendix:keywords}
In this appendix we present the hashtags used to construct our endorsement networks. In the tables below, we omit the \#, but note that all keywords were supplied to Twitter's API with the preceding \#. These hashtags have been translated into English with an eye toward ease of understanding. Please contact us for the original Finnish version. Slogans or catchphrases are marked with *.

\singlespacing
\subsection{Climate}
\begin{multicols}{3}
\small
\begin{verbatim}
carbon sink
carbon storage
climate
climate strike
climate change
climate politics
climate elections
nature conservation
biodiversity
now we must*
show you have a spine*
swamp protection
mass extinction wave
peat
the environment 
\end{verbatim}
\end{multicols}

\subsection{Issues}

\subsubsection*{Immigration}
\begin{multicols}{3}
\small
\begin{verbatim}
some border
integration of immigrants
child rape
immigrant
immigration
immigration policy
immigration election
intruder
immigrant
intruder
multiculture
multiculturalism
muslim
forced returns
paperless immigrants
close the borders
racism
racists
refugees
sex crimes
immigrants
stop grooming
Finland is racist
tolerance 
asylum seeker
asylum seekers
asylum seeker flood
asylum
\end{verbatim}
\end{multicols}

\subsubsection*{Social Security}
\begin{multicols}{2}
\small
\begin{verbatim}
homelessness
pension
inequality
dependency ratio
welfare state
sustainability deficit
poverty
local government reform
basic income
social security
social and healthcare system
social and healthcare system reform
health care business
employment
unemployment security
active model*
\end{verbatim}
\end{multicols}

\newpage
\subsubsection*{Economic Policy}
\begin{multicols}{2}
\small
\begin{verbatim}
economy
economic policy
taxation
export industry
\end{verbatim}
\end{multicols}

\subsubsection*{Education}
\begin{multicols}{2}
\small
\begin{verbatim}
students in applied science
research by students in applied science
vocational training
more than high school
no cuts in education
education
education promise
education is key
education elections
international students
student benefits
this is why Science is important
\end{verbatim}
\end{multicols}

\subsection{Parties}
\subsubsection*{Social Democratic Party}
\begin{multicols}{2}
\small
\begin{verbatim}
social democrats
sdp
same direction*
future line*
\end{verbatim}
\end{multicols}

\subsubsection*{Finns Party}
\begin{multicols}{2}
\small
\begin{verbatim}
finns party
finns party helsinki
finns party members and supporters
this is why true finns*
another big bang coming*
\end{verbatim}
\end{multicols}

\subsubsection*{National Coalition}
\begin{multicols}{2}
\small
\begin{verbatim}
national coalition party
national coalition party cruise
national coalition party cruise 2019
we trust in finland*
at right*
\end{verbatim}
\end{multicols}

\subsubsection*{Centre Party}
\begin{multicols}{2}
\small
\begin{verbatim}
centre party
\end{verbatim}
\end{multicols}

\subsubsection*{Green League}
\begin{multicols}{3}
\small
\begin{verbatim}
sensible green
the greens party
show your true nature*
\end{verbatim}
\end{multicols}

\subsubsection*{Left Alliance}
\begin{multicols}{2}
\small
\begin{verbatim}
the left
the left alliance party
the party council of the left alliance
\end{verbatim}
\end{multicols}
\end{document}